
\hfuzz50pt
\input epsf           


\parindent=0pt
\parskip=3pt plus 1pt
\font\titel=cmbx10 scaled \magstep2
\bigskip
\centerline{{\titel Self--gravitating fluid shells and their}}
\vskip .5cm
\centerline{{\titel non--spherical oscillations in Newtonian theory}}
\vskip2.5cm
\centerline{\titel{Ji\v r{\'\i} Bi\v c\'ak$^{1,2}$  and Bernd G. Schmidt$^1$}}
\vskip 2cm
\centerline{$^1$ Albert Einstein Institute}

\centerline{Max--Planck Institute for Gravitational Physics}

\centerline{Schlaatzweg 1}

\centerline{14473 Potsdam, Germany}

\vskip 1cm
\centerline{$^2$ Permanent address:}

\centerline{Department of Theoretical Physics}

\centerline{Faculty of Mathematics and Physics}

\centerline{Charles University}

\centerline{V Hole\v sovi\v ck\'ach 2}

\centerline{180 00 Prague 8, Czech Republic}
\vfill\eject

{\bf  Abstract }
\vskip 1cm
We summarize the general formalism describing surface flows in
three--dimensional space in a form which is suitable for various astrophysical
applications. We  then apply the formalism to the analysis of non--radial
perturbations of self--gravitating spherical fluid shells.

Spherically symmetric gravitating shells (or bubbles) have been used in
numerous model problems especially in general relativity and cosmology. A
radially oscillating shell was recently suggested as a model for a variable
cosmic object. Within Newtonian gravity we show that self--gravitating
static fluid shells are unstable with respect to linear non--radial
perturbations.  Only shells (bubbles) with a negative mass (or with a charge
the repulsion of which is compensated by a tension) are stable.

\vskip 2cm
{\bf\it{ Subject headings:}} gravitation --- hydrodynamics --- instabilities ---
 stars: oscillations --- supernovae: general

\vfill\eject

  {\bf  1. Introduction }
\vskip 1cm
It is interesting to see how modelling problems by thin shells whose thickness
is being ignored is employed in so many different sciences as general
relativity, astrophysics, cosmology, elasticity or chemical engineering. In
general relativity thin spherical shells of dust or perfect fluids
\footnote{$^3$}{By "surface perfect fluids" we mean the surface distributions of matter
with isotropic stress distribution tangent to the surface.} have
frequently been used to analyse basic issues of gravitational collapse, in both
its classical and quantum aspects (see e.g. Barrab\`es \& Israel 1991; Friedmann,
Louko \& Winters--Hilt 1997, and references therein); in astrophysics expanding
spherical shells model supernovas  (e.g. Vishniac 1983;  Sato \& Yamada 1991); the
chief motivation to study shells in cosmology has been observations of bubble--like
structures in the distribution of galaxies  (e.g. Peebles 1993;   Turok 1997) but also
the physics of the early universe in which a region of false vacuum is separated by a
domain wall (modelled usually as a spherical shell) from a region of true vacuum (e.g.
Blau, Guendelman \& Guth 1987; Berezin, Kuzmin \& Tkachev 1987; Kolitch \& Eardley
1997); in elasticity the theory of rods and thin shells goes back to the last century (cf.
Love 1944), and in chemical engineering the mathematical description of the dynamics of an
interface is important in such problems as calming of water waves by oil or in
distillation and liquid extraction (Scriven 1960). A theoretical physicist of the new age
would of course add membranes (or rather $D$--branes) moving in higher--dimensional
spacetimes in superstring theories.

We, as relativists, worked on various problems connected with thin shells (e.g. Bi\v c\'ak \&
Ledvinka 1993; H\'aj\'\i\v{c}ek
\& Bi\v c\'ak 1997). Recently, one of us investigated
non--radial oscillations of static self--gravitating spherical fluid shells in general
relativity and their Newtonian limit (Schmidt 1998). However, we could not find a
reference to this problem solved within Newtonian theory. Radial oscillations of spherical
shells in which gravity is balanced by the  surface pressure were analyzed  both in the
Newtonian and relativistic case by several authors, even for shells surrounding a compact
object (Brady, Louko \& Poisson 1991, and references therein). Most recently, in this
journal such radially oscillating shells have been suggested as a model for variable cosmic
objects (N\'u\~nez 1997). Non--radial oscillations are more difficult and even in Newtonian
theory require some differential geometry because, for example, the correct form of the
equation of continuity for a surface flow depends on the second fundamental form of an
embedded surface in $R^3$.

In this work we investigate the non--radial oscillations of self--gravitating (or
charged) spherical shells in Newtonian theory in detail. In Section 2 we review the
formalism needed to describe surface flows. We here essentially follow the exposition
given by Aris (1989) in  chapter 10 of his book which, in turn, ''is in the nature a
somewhat extended gloss" on a paper by Scriven (1960) (both works thus emanating from a
chemical engineering department). Our discussion is, of course, much shorter, however, it
generalizes both mentioned works in two respects. We define a ''coordinate
system fixed in the surface moving in space"
 and we show how the equation of motion and the continuity equation get modified if other
(general) coordinates are used within the surface since in concrete problems the
''fixed" (Gaussian) coordinates are not practical at all. Secondly,
although we use index notation we make occasionally contact with the formulation of
 mathematical elasticity theory by Marsden \& Hughes (1983) which is based on
the index--free formulation of modern differential geometry. In fact, Marsden and Hughes'
text also touches on shell theory but it does not give the equations of motion and
whenever it uses coordinates these are again only the ''Gaussian--type"
 coordinates.\footnote{$^4$}{In Marsden \& Hughes (1983), the comprehensive work by Naghdi
(1972) is quoted as the standard reference for shells. As much as this work may be
preferable for dealing with many aspects of elasticity problems, for our
purposes, we found Aris (1989) more useful.}

In Section 3 we first discuss the general dynamics of a self--gravitating shell which is
topologically spherical but may largely deviate from a sphere. We believe that Section 2
and the beginning of Section 3 may serve as a basic formalism for analyzing, for example,
expanding or collapsing non--spherical self--gravitating shells in Newtonian theory in
astrophysically realistic situations. In the second part of Section 3 we derive the
conditions for a spherical shell to be in equilibrium.

Section 4 is devoted to the derivation of the equations of motion and the continuity
equation for linear perturbations of the static solution obtained in Section 3.

In Section 5, the stability  of the
static solution is analyzed. We prove that although, with an appropriate equation of state,
the shell is stable with respect to radial oscillations, it is unstable if it is perturbed
non--radially. It can thus hardly serve as a model for variable cosmic objects as
suggested recently (N\`u\~nez 1997).

At the end we notice the fictitious, but amusing case of shells with negative gravitational
and inertial mass.  Since the time of an interesting work by Bondi (1957) it is well--known
that in principle a negative mass can exist in the sense that it is not forbidden by
classical physics. Some of its amusing properties were recently described by Price
(1993). We show that spherical static shells with negative mass are, in fact, stable with
respect to non--radial oscillations! Although there is no evidence that a negative mass
exists in the real universe, in numerical relativity spacetimes containing negative mass
 solutions of Einstein's equations serve as  testbeds.

Finally, by considering formally the gravitational constant to be negative, we show that
charged shells, in which  the repulsive effects of  the charges is compensated by a
tension, are stable. We also give   intuitive physical arguments for the results of the
stability analysis in all three cases.
\vskip 2cm
{\bf  2. Equations of motion for surface fluids}
\vskip 1cm
The flow in a surface is more complicated than an infinitely extended
3--dimensional flow because the 2--dimensional surface (shell) can move in the
3--dimensional space which surrounds it.

Let $(t, x^i)$ be inertial coordinates
 in Newtonian spacetime. The metric $g_{ik}$ is the time--independent
metric on the flat Euclidian 3--space $R^3$ (depending on
the application one may use Cartesian,  polar or other coordinates).

Functions $x^i=\hat x^i(t,a^\alpha),$ $\alpha=1,2$, decribe the world lines of the
particles of the fluid; $a^\alpha$ are thus comoving (Lagrangian) coordinates. We assume
that for fixed times
$t$ the points
$\hat x^i(t,a^\alpha)$ form a 2--surface $\Sigma_t$ in Euclidian space.  We may
think of the shell  as a 3--surface in 4--dimensional spacetime  which is formed by the flow
lines, or as a sequence of 2--surfaces $\Sigma_t$ in $R^3$.

The space component of the
tangent vector to the curves in 4--space, i.e. the velocity of a particle of the fluid in
$R^3$ is
$$
U^i=\left({\partial\hat x^i\over \partial t}\right )\!\!_a\ ,
\eqno(1)
$$
and its acceleration
$$
A^i=\left({\partial^2\hat x^i\over \partial t^2}\right)\!\!_a\ .
\eqno(2)
$$
We are free to use arbitrary coordinates $z^\alpha$ on $\Sigma_t$. (Objects
intrinsic to $\Sigma_t$ will have Greek indices; in $R^3$ we use Latin indices.)
In particular we will use coordinates $y^\alpha$, obeying  the condition that
the velocities of the points with $y^\alpha$=const are orthogonal to $\Sigma_t$ for any
$t$. In these coordinates we describe the shell by
$$
x^i=f^i(t,y^\alpha)\ .
\eqno(3)
$$
Then the vectors in $R^3$ given by
$$
t^i_\alpha=\left({\partial f^i\over \partial y^\alpha}\right)\!\!_t
\eqno(4)
$$
are tangent to $\Sigma_t$ and the velocities of the points $y^\alpha$=const,
$(\partial f^i/\partial t)_y$, in $R^3$ are perpendicular to  $t^i_\alpha$:
$$
g_{ij}t^i_\alpha \left({\partial f^j\over\partial t}\right)\!\!_y=0 \ ,
\eqno(5)
$$
where $g_{ij}$ is the metric in  $R^3$. These coordinates can be constructed by drawing
the orthogonal curves to the family of 2--surfaces $\Sigma_t$ in $R^3$. Choosing some
coordinates $y^\alpha$ on one surface and taking $y^\alpha$ constant along the orthogonal
congruence defines the coordinates $y^\alpha$. These coordinates are unique up to a
transformation $y^{\alpha'}(y^\beta)$, independent of $t$. (In general relativity such a
coordinate system is called ''with vanishing shift" --- see, e.g., Wald 1984.)

In general, it may be more convenient  to use other coordinates in
$\Sigma_t$, say $z^\alpha$, and describe the moving  $\Sigma_t$ in $R^3$ by
$$
x^i=\zeta^i(t,z^\alpha),\ \ \ \tau^i_\alpha=\left({\partial\zeta^i\over\partial
z^\alpha}\right)\!\!_t\ ,
\eqno(6)
$$
although $(\partial\zeta^i/\partial t)_z$ is not perpendicular to $\tau^i_\alpha$ ---
imagine, for example, a motion of the spherical surface into a highly oblate ellipsoidal
surface which is described by
$$
r=R(t,\theta,\varphi)
\eqno(7)
$$
in the standard spherical coordinates in $R^3$, with $z^\alpha=(\theta,\varphi)$.

We assume $y^\alpha=\hat y^\alpha(t,a^\beta)$ and, inversely, $a^\beta=\hat
a^\beta(t,y^\alpha)$; the same for $z^\alpha$. The 2--dimensional metric,
$h_{\alpha\beta}(t,y^\gamma)$,  determines the line element in $\Sigma_t$,
$$
dl^2=h_{\alpha\beta}dy^\alpha dy^\beta\ ,
\eqno(8)
$$
which can be considered as induced by  (pull--back of ) the metric $g_{ij}$ of $\Sigma_t$,
$$
h_{\alpha\beta}={\partial f^i\over \partial y^\alpha}{\partial f^j\over \partial y^\beta}
\ g_{ij}\ ,
\eqno(9)
$$
similarly for $z^\alpha$.
The surface velocity of the fluid is defined by
$$
V^\alpha=\left({\partial\hat y^\alpha\over \partial t}\right)\!\!_a\ ,
\eqno(10)
$$
and the acceleration by
$$
A^\alpha=\left(\partial V^\alpha\over\partial t\right)\!\!_y + V^\alpha{}_{|\beta} V^\beta\ ,
\eqno(11)
$$

where the vertical bar denotes the covariant derivative with respect to $h_{\alpha\beta}$.
 Such quantities defined analogously in general surface coordinates $z^\alpha$ do not have
a natural geometrical (physical) meaning as in the expressions (10) and (11).

In order  to see this, imagine a particle of the fluid with fixed $a^\alpha$ moving in
$R^3$ according to $x^i=\hat x^i(t,a^\alpha)$. Its velocity $U^i$ and acceleration $A^i$
are given by expressions (1) and (2). Using the coordinates $y^\alpha$, these may be
written as

$$
U^i={\partial f^i\over\partial t}+t^i_\alpha V^\alpha\ ,
\eqno(12)
$$

and

$$
A^i={\partial U^i\over\partial t}+U^i_{|\alpha} V^\alpha\ ,
\eqno(13)
$$

where $t^i_\alpha, V^\alpha$ are given by  (4) and (10), and the covariant derivative
is defined by $U^i_{|\alpha}=U^i_{,\alpha} + \Gamma ^i_{jk} V^j\ t^k_\alpha$.
Let $n^i$ be the unit normal to $\Sigma_t$. Regarding equations (12) and (5), we find

$$
t^i_\alpha U_i=V_\alpha\ ,\ \ \ \  n_iU^i=n_i{\partial f^i\over\partial t}\ ,
\eqno(14)
$$
i.e.,  equation (12) represents the decomposition of the velocity in $R^3$ into its
normal and tangential parts with resepect to $\Sigma_t$. (In geometrical language,
$V_\alpha$ is the pull--back of $U_i$ to $\Sigma_t$.) It is easy to see that one can also
write
$$
A^i=(n_j A^j)n^i+t^i_\alpha A^\alpha\ ,
\eqno(15)
$$
where $A^\alpha$ is given by expression (11) so that it represents the particle's
acceleration along $\Sigma_t$.
Now using general coordinates $z^\alpha$ for which equation (5) is {\it not} satisfied,
we can still write
$$
\tilde V^\alpha=\left({\partial \hat z^\alpha\over\partial t}\right)\!\!_a\ ,
\eqno(16)
$$
for $\tilde A^\alpha$ analogously with equation (11), and we obtain $U^i$ in the form
(see  eq.(6))
$$
U^i={\partial\zeta^i\over\partial t}+\tau^i_\alpha\tilde V^\alpha\ .
\eqno(17)
$$
However, we now get
$$
\tau^i_\alpha U_i= \tau^i_\alpha\ g_{ij}\left({\partial\zeta^j\over\partial t}\right)\!_z+
\tilde V_\alpha\ ,
\eqno(18)
$$
so that $\tilde V_\alpha$ is not a total projection of $U^i$ on $\Sigma_t$ (and neither
its pull--back) because $(\partial\zeta^i/\partial t)_z$ has also a non--vanishing
projection onto $\Sigma_t$. When using such coordinates we just have to remember that the
surface velocity of the fluid is given by the whole r.h.s. of equation (18).

Before considering the dynamics let us point out that $h_{\alpha\beta}$ and
$k_{\alpha\beta}$ are geometrical quantities depending only on the position of the surface
$x^i=\zeta^i(t,z^\alpha)$ at a given time, but independent of the particles flow. In fact, the
set of surfaces define intrinsically the normal velocity field $U_\perp(t,z^\alpha)$ at
each point of the surface at a given time. The explicit expression for this field can be
given in terms of our Gaussian coordinates by $ \left({\partial f^i\over \partial
t}\right)_y=U_\perp n^i$. After introducing particles their $U^i$ can be decomposed as
$U^i= n^i U_\perp + V^i_{tan}$ and $V_\alpha$ can be determined intrinsically by $V_\alpha
= (V_{tan})_it^i_\alpha$.

Now let $F(t,y^\alpha)$ be any function defined on $\Sigma_t$. Denoting by $S_t$ any part
of $\Sigma_t$ then one can derive the following analogue of Reynold's transport theorem
for the material (or convective) derivative of the integral of $F$ over $S_t$:

$$
{d\over dt}\int_{S_t} \ FdS= \int_{S_t} \left[
\left({\partial F\over\partial t}\right)_{\!a}+ F\
\left({\partial\over\partial t}\ln\sqrt h\right)_{\!\!a}\
\right] dS\ $$
$$=
\int_{S_t} \left[
\left({\partial F\over\partial t}\right)_y+ F\  {\dot h\over 2h}
+(FV^\alpha)_{|\alpha}
\right] dS\ ,
\eqno(19)
$$
where $h=$det$(h_{\alpha\beta})$ and the dot means $(\partial/\partial t)_y$. In
particular, let $\sigma(y^\alpha,t)$ be the surface density of the fluid. Then, if the
mass of any part of $\Sigma _t$ is conserved, the transport theorem (19) implies the
continuity equation
$$
{d\sigma\over dt}+\sigma
V^\alpha{}_{|\alpha}+\sigma {{\dot h\over 2 h}}=0\ ,
\eqno(20)
$$
where $d\sigma/dt=(\partial\sigma/\partial t)_y+ V^\alpha\partial\sigma/\partial y^\alpha$
is the material derivative of $\sigma$.

Let us define the  external curvature tensor $k_{\alpha\beta}$ of the surface by
$$
t^i_{\alpha|\beta}=-k_{\alpha\beta}n^i\ ,
\eqno(21)
$$
where $n^i$ is the unit normal to $\Sigma_t$ as before, and the mean curvature is defined
by\footnote{${^5}$}{Our definitions agree with those in Marsden  \& Hughes (1983), but not
with Aris (1989):
$b_{\alpha\beta}=-k_{\alpha\beta}$, $H({\rm Aris)}=-{1\over 2} H$.}
$$
H={\rm  tr }k_{\alpha\beta}=h^{\alpha\beta}k_{\alpha\beta}\ .
\eqno(22)
$$
A short calculation using standard geometry (see, e.g., exercise 10.41 in Aris 1989) shows
that the last term in equation (20) can be written as
$$
{\dot h\over 2 h}=H n^jU_j\ ,
\eqno(23)
$$
where $U^j$ is the space velocity of a fluid particle given by equation (12).  The
continuity equation (20) can thus be rewritten in the form
$$
{d\sigma\over d t}+\sigma V^\alpha{}_{|\alpha}+\sigma H n_jU^j =0\ ,
\eqno(24)
$$
which is the form given by Marsden \& Hughes (1983) in the Theorem 5.15 and written in the
Box 5.2 in the Gaussian coordinate system attached to $\Sigma_t$ so that equation (5) is
satisfied. In general coordinates $z^\alpha$ on $\Sigma_t$, we still obtain the
continuity equation in the form (24) if, instead of $V^\alpha$ , we substitute
$g^{\alpha\beta}\tau^i_\beta U_i$ given in equation (18); only the  expression
$g^{\alpha\beta}\tau^i_\beta U_i$ is the component of $U^i$ parallel to $\Sigma_t$
(geometrically the pull--back of $U_i$ on
$\Sigma_t$).

In order to derive equations of motions for the shell one starts from balancing the rate
of change of  momentum of a portion of the shell with the total force acting on it.
 Let the properties of the fluid be described
by the surface stress tensor $T^{\alpha\beta}$. For example, in the case of a Newtonian
 surface fluid in which the viscous  stress depends linearly on the rate of strain, the
stress tensor reads (Aris 1989)
$$
T^{\alpha\beta}=-p g^{\alpha\beta} + \kappa S^\sigma_\sigma g^{\alpha\beta}
+\epsilon E^{\alpha\beta\rho\sigma}S_{\rho\sigma}\ ,
\eqno(25)
$$
where $p$ is the surface pressure, the surface deformation tensor is
$S_{\rho\sigma}={1\over 2} \dot g_{\sigma\rho}+{1\over 2}(V_{\rho |\sigma} +V_{\sigma
|\rho})$,
$E^{\alpha\beta}{}_{\lambda\mu}=
\delta^\alpha_\lambda\delta^\beta_\mu+\delta^\alpha_\mu\delta^\beta_\lambda
-g^{\alpha\beta} g_{\lambda\mu}$, and $\kappa$ and $\epsilon$ are the coefficients of
dilatational and shear surface viscosity. We wrote down expression (25) just for
illustration, in the following we shall consider only ideal surface fluids, i.e.,
$\kappa=\epsilon=0$, but at the moment we leave a general $T^{\alpha\beta}$.

Let us now assume that, besides the internal pressure, there acts a  surface force,
$F^\alpha$, per unit area of the fluid. We require the balance of momentum in the direction
of an arbitrary smooth covariantly constant vector field $C^\alpha$ in the form
$$
{d\over dt}\int_{S_t} \sigma V^\alpha C_\alpha dS=\int_{S_t} F^\alpha C_\alpha dS
+\int_{\partial S_t} T^{\alpha\beta}\nu_\beta C_\alpha dl\ ,
\eqno(26)
$$
where
$$
T^\alpha dl \equiv T^{\alpha\beta}\nu_\beta dl
\eqno(27)
$$
 is the
surface stress vector  acting on a linear element $dl$ (in $\partial S_t$ in $\Sigma_t$) with
a unit normal
$\nu^\alpha$. Converting the last integral to a surface integral by Green's theorem, and
using the transport theorem (19) and the continuity equation (20) on the left--hand
side, we obtain the intrinsic (surface) equations of motion
$$
\sigma A^\alpha = T^{\alpha\beta}{}_{|\beta} + F^\alpha\ ,
\eqno(28)
$$
where $A^\alpha$ is given by equation  (11). (As usual, the balance of angular momentum
holds since
$T^{\alpha\beta}=T^{\beta\alpha}$.)

Finally, consider the motion in $R^3$. The external force, $F^i$, will in general have a
component normal to the surface $\Sigma_t$,
$$
F^i=(n_jF^j) n^i + t^i_\alpha F^\alpha\ ,
\eqno(29)
$$
 as the acceleration given by  equation  (15). Notice that $t^i_\alpha F^\alpha$, and
similarly $t^i_\alpha T^\alpha dl$ (see eq. (27)), are just space components of the
surface external force and surface stress (in geometrical language they are the
push--forward vectors of $F^\alpha$ and $T^\alpha dl$).
Starting from the balance of momentum in $R^3$ in the direction of an arbitary smooth
covariantly constant vector field $K^i$ analogously to equation (26) (now with
quantities
$\sigma U^i, F^i$ and $t^i_\alpha T^{\alpha\beta}\nu_\beta$), we find the complete
3--dimensional form of the equations of motion:
$$
\sigma A^i=(t^i_\alpha T^{\alpha\beta})_{|\beta} + F^i\ .
\eqno(30)
$$
Expressing $t^i_{\alpha|\beta}$ by using equation (21)  we can write them as
$$
\sigma A^i=t^i_\alpha T^{\alpha\beta}{}_{|\beta}-k_{\alpha\beta} T^{\alpha\beta} n^i +
F^i\ .
\eqno(31)
$$
In the case of a perfect fluid the equations of motion become
$$
\sigma {dU^i\over dt}=-t^i_\alpha g^{\alpha\beta}p_{|\beta}+
p H  n^i + F^i\ ,
\eqno(32)
$$
where $H$ is the mean curvature (22).

It is instructive to project these equations into the directions tangent and normal to
$\Sigma_t$. The tangential part is given by equations (28), the normal part becomes
$$
\sigma (n_jA^j)= p H + n_jF^j\ ,
\eqno(33)
$$
which demonstrates how the surface pressure influences the motion in the direction
normal to the surface if the surface is bent in $R^3$ so that it has non--zero external
curvature.

The equations of motion (32) and the continuity equation (24) determine the motion
of the fluid.
In the Lagrangian approach the velocity and acceleration are given by expressions (1),
(2), all other quantities are also functions of $(t, a^\alpha)$, and one seeks
solutions $\hat x^i(t, a^\alpha)$, $\sigma(t,a^\alpha)$, assuming some equation of state
$p=p(\sigma)$ and external force $F^i$ are given.
In the Eulerian description one solves for $f^i(t,y^\alpha)$ and $\sigma(t,y^\alpha)$ in
terms of which
$U^i$, $t^i_\alpha$, $V_\alpha$, $g_{\alpha\beta}$ are determined by equations (4), (9),
(10), (12). Let us recall yet that we need not  use  ``Gaussian--type" coordinates
$y^\alpha$ attached to the surface so that equation (5) is satisfied. We can solve for
functions
$\zeta^i(t,z^\alpha)$  (see eq. (6)), we only have to remember that the surface velocity
of the fluid (as it appears, for example, in the continuity equation (24)) is given by
the whole right--hand side of equation (18).
\vskip 2cm
{\bf  3. The self--gravitating shell }
\vskip 1cm
The gravitational field determined by a surface distribution of matter has a
potential $\Phi$ which is continuous at the surface (see, e.g., Kellog 1967). Assume
that the shell $\Sigma_t$ is topologically a sphere. The derivatives of $\Phi$ have
limits on both sides of the shell, and the (covariant component of) gravitational
force at a point of the shell is given by the mean
$$
F_i=-{1\over 2}\sigma(^+\!\Phi_{,i}+^-\!\Phi_{,i})\ ,
\eqno(34)
$$
where $^+\!\Phi\ ,^-\!\Phi$ are the potentials on the two sides of $\Sigma$ and   commas
denote partial derivatives (see, e.g., Purcell 1965 for a derivation of equation
(34) in the analogous electric case).

Hence, if the shell of perfect fluid moves under its own gravitational field, the
continuity equation has the form (24), and the equations of motion (32) become
$$
\sigma {dU^i\over dt}=-t^i_\alpha g^{\alpha\beta}p_{|\beta}+
p H  n^i -{1\over 2}\sigma g^{ij}(^+\!\Phi_{,j}+^-\!\Phi_{,j})\ .
\eqno(35)
$$
In order that these equations  indeed determine the motion of the shell we need
to know $\Phi$ in terms of the shell's variables.

Let us  assume that, in general, the position of the shell is given in spherical
coordinates by $r=R(t,\theta,\varphi)$, as in equation (7). Then we want to solve the
Poisson equation
$$
{1\over r}{\partial^2\over\partial r^2}(r\Phi)+{1\over r^2\sin\theta}{\partial\over
\partial\theta}(\sin\theta{\partial\Phi\over\partial\theta})+{1\over r^2\sin^2\theta}
{\partial^2\Phi\over\partial\varphi^2}
=4\pi G \rho(t,r,\theta,\varphi)\ ,
\eqno(36)
$$
where the matter density is non--vanishing only on $\Sigma_t$. Introducing the
surface matter density $\sigma(t,\theta,\varphi)$, we find
$$
\rho(t,r,\theta,\varphi)=\sigma(t,\theta,\varphi){\sqrt{h(t,\theta,\varphi)}
\over R^2(t,\theta,\varphi)\sin\theta}\ \delta (r-R(t,\theta,\varphi))\ ,
\eqno(37)
$$
where $\delta$ is the Dirac delta  and $h=$det($h_{\alpha\beta}$), with
$h_{\alpha\beta}$ being the metric on $\Sigma_t$ induced by the spatial metric
$g_{ij}$ as in equation (9). Equation  (37) follows from the general relations for
3--dimensional distribution $\delta_F$ with support on  a 2--surface $\Sigma$, given by
$F(r,\theta,\varphi)=0$, which for any nice function $f(r,\theta,\varphi)$ requires
$$
\int_{R^3}f(r,\theta,\varphi)\,\delta_F \,r^2 \sin\theta drd\theta d\varphi=
\int_\Sigma f|_\Sigma \ d\Sigma\ ,
\eqno(38)
$$
where $f|_\Sigma $ is the restriction of $f$ to $\Sigma$. We shall return to the
solution of equation (36) in the next section. Now we consider the simplest case --- that
of spherical symmetry.

We thus assume $\sigma, p$ independent of $\theta,\varphi$; $U^i=(\dot R,0,0)$, and the
mean curvature of a sphere of radius $r=R(t)$ is
$$
H={2\over R}\ ,
\eqno(39)
$$
the normal $n^i=(1,0,0)$, the potential vanishes inside and reads
$$
^+\!\Phi=-{GM\over r}
\eqno(40)
$$
outside the shell. The equations of motion (35) thus reduce to
$$
\sigma \ddot R = {2p\over R} - {1\over 2}\sigma {GM\over R^2}\ ,
\eqno(41)
$$
and the continuity equation (24) is
$$
{d\sigma\over dt}+{2\sigma\dot R\over R} =0\ .
\eqno(42)
$$
The continuity equation can be immediately integrated to yield
$$
\sigma R^2=\sigma_0 R^2_0\ ,
\eqno(43)
$$
where the right--hand side denotes values at a fixed time (say $t=0$). It is evident
that equation (43) means the conservation of mass; it is also instructive to see how the
outward directed radial force due to the surface pressure, i.e. the term $2p/R$
 in equation (41), can be derived from elementary considerations of the force acting on
a surface element $dS=R^2\sin\theta d\theta d\varphi$ from its surrounding.

Substituting $M=4\pi\sigma R^2$, an equation of state $p=p(\sigma)$, and eliminating
$\sigma$, by equation  (43), from the equation of motion (41), we  can solve equation
(41) for the radius of the shell $R(t)$. We shall discuss radial oscillations in the next
section. Now we just notice that equation  (41) admits a unique static solution for a given
$M$ and $R=\bar R$. In this static case the pressure is
$$
\bar p ={1\over 4}{G\bar\sigma M\over \bar R}=\pi G\bar\sigma^2={1\over
16\pi}{G M^2\over \bar R^3}\ .
\eqno(44)
$$
The positive surface pressure is needed to balance the inward directed
gravitational force.

It is amusing to observe that exactly the same static situation
arises if $M<0,\bar\sigma<0$.
How is it possible that a {\it negative} gravitational mass which {\it repels} all
masses and thus pushes the elements of the shell outwards is compensated by a {\it
positive} pressure which, as we saw above, exerts, apparently, an outward directed
radial force on each element? The resolution of this paradox comes from the fact
that we assumed both gravitational and inertial mass of the shell to be negative. A
non--gravitational force (as the pressure) acting in a given direction on a negative
inertial mass gives it an acceleration in the opposite direction (just by ${\bf
F}=m{\bf a})$! So the positive pressure, in fact, accelerates the elements of the shell in
the inward direction and is just compensated by the repulsive gravitational action of
the negative gravitational mass.
We shall see how these effects can influence the stability of the static shell in
the following sections.

Another alternative is to change the sign of $G$. Gravity becomes then electricity
in which the charges of the same sign repel each other. Equation (44) says that we need
a negative $\bar p$, a tension as in soap bubbles, to balance such a shell.
\vskip 2cm
{\bf  4. The linearized equations of motion }
\vskip 1cm
In general the solution of the coupled system of nonlinear equations
 (24), (35) and (36)
describing the motion of a highly deformed shell is complicated.
Our goal here is to investigate
linearized perturbations (oscillations) of a static spherical shell
satisfying the conditions (44). Since the background solution is
spherically symmetric we can, without loss of generality, assume
that the perturbations are axisymmetric, i.e., independent of the
azimuthal coordinate $\varphi$. The position $\Sigma_t$ of the shell at
time $t$ is thus given by
$$
r=R(t,\theta)\ ,
\eqno(45)
$$
with $t$ fixed. If we let $\theta$ and $t$ change,  equation (45) describes a
3--surface in (Newtonian) spacetime. Now the fluid can move in the shell;
denote its surface velocity by
$$
\tilde V^\theta(t,\theta)=W(t,\theta)\ .
\eqno(46)
$$
The vector
$$
\tau^i_\theta={\partial x^i\over\partial\theta}=(\ R_{,\theta},\ 1,\ 0)
\eqno(47)
$$
is tangent to $\Sigma_t$. The fluid velocity in space, given by
 $U^i=(\partial x^i/\partial t)_\theta +\tau^i_\theta\tilde V^\theta $
(cf. eqs. (6 ), (17)), reads

$$
U^i(t,\theta)=(\dot R+R_{,\theta} W,\ W,\ 0)\ ,
\eqno(48)
$$
where $\dot R=\partial R/\partial t$. The surface metric
$^{(2)}\!g_{\alpha\beta}$ induced on $\Sigma_t$ is
$^{(2)}\!g_{\theta\theta}=R^2_{,\theta} +R^2$,
$^{(2)}\!g_{\varphi\varphi}=R^2\sin^2\theta $, so that the covariant component
$\tilde V_\theta=(R^2_{,\theta} +R^2)W$; hence,
$$
\tau^i_\theta\ U_i=R_{,\theta}(\dot R +R_{,\theta}W)+ R^2 W\ ,
\eqno(49)
$$
and equation (18) is indeed satisfied.

The acceleration appearing on the left--hand side of equation (35) is (see eq.
(13))
$$
{dU^i\over dt}={\partial U^i\over\partial t}+U^i_{|\alpha}\tilde V^\alpha
={\partial U^i\over\partial t}+\left({\partial U^i\over\partial\theta}+\Gamma^i_{jk}
U^j\tau^k_\theta\right)\tilde V^\theta\ ,
\eqno(50)
$$
which in spherical coordinates implies
$$
{dU^r\over dt}=\ddot R+ 2\dot R_{,\theta} W +R_{,\theta}\dot
W+W(R_{,\theta}W)_{,\theta}-RW^2\ ,
\eqno(51)
$$
$$
{dU^\theta\over dt}=\dot W +W{\dot R\over R}+W
W_{,\theta}+2W^2{R_{,\theta}\over R}\ .
\eqno(52)
$$
(Since $\tilde V^\theta, U^\theta$ are coordinate components, their dimension and
thus the dimension of $W$ is s$^{-1}$, whereas $U^r$ has the usual dimension cm
s$^{-1}$ in CGS units.)
The normal to the shell is given by
$$
n_i={1\over\sqrt{1+({R_{,\theta}/ R})^2}}(1\ ,
 -R_{,\theta}\ ,\; 0)\ ,
\eqno(53)
$$
$$
n^i={1\over\sqrt{1+({R_{,\theta}/ R})^2}}(1\ ,\ -R^{-2}R_{,\theta}\ ,\; 0)\ ,
\eqno(54)
$$
and the mean curvature turns out to be
$$
H={1\over R\sqrt{1+({R_{,\theta}/ R})^2}}\left[
2-{1\over R}(R_{,\theta}\cot\theta +R_{,\theta\theta})
+{1\over {1+({R_{,\theta}/R})^2}}\left({R_{,\theta}\over
R}\right)^2\left({R_{,\theta}\over R}\right)_{,\theta}
\right]\ .
\eqno(55)
$$
We can write down the exact form of the continuity equation (24) in the
general case by substituting for $n_j$, $U^j$ and $H$ the expressions above
and for $V^\alpha$ the expression (49) since the surface coordinates
$\theta$, $\varphi$ are generalized coordinates as $z^\alpha $ in Section 2
rather than $y^\alpha$ (the lines ($\theta,\varphi)=$const are not
perpendicular to the surface $r=R(t,\theta$)). However, we shall not analyze
the general case further, we shall now linearize both the continuity equation
and the equations of motion around the static solution satisfying equation (44).

To derive the linearized equations we consider  1--parameter
families $R(t,\theta,\epsilon)$,
$W(t,\theta,\epsilon)$ of  shell solutions. The coordinates in the shell and in
the embedding space are uniquely fixed. Hence we obtain  a description of the
linearized equations in a particular (coordinate) gauge. As always we assume that the
family is smooth in $\epsilon$ and that we can interchange
$\epsilon$--derivatives and spacetime derivatives.

Let $R(t,\theta,0)=\bar R$, $W(t,\theta,0)=0$ be a static shell.  We denote
background quantities with an overbar and  the perturbation of a quantity $Q$
by $\delta Q$. It is easy to see that in the linearized case the coordinates
become Gaussian in linear order, i.e. as coordinates $y^\alpha$ used in
Section 2; indeed neglecting higher--order terms, equation (5) is satisfied.
The linearized acceleration (51), (52) is
$$
\delta A^i=(\ \delta\ddot R,\ \dot W,\ 0)
\eqno(56)
$$
because the background acceleration vanishes. We have now to determine the
radial and tangential components of the linearization of all terms in equations  (35).
For the inner forces due to the pressure gradient we obtain only a tangential
component
$$
-\delta(t^i_\alpha g^{\alpha\beta}p_{,\beta})=-t^i_\alpha g^{\alpha\beta}\delta
p_{,\beta}=-{1\over\bar R^2}\delta^i_\theta\delta p_{,\theta}
\eqno(57)
$$
because the background pressure is independent of $\theta$.
The perturbation of the normal force is
$$
\delta(pHn^i)=\delta p \bar H\bar n^i +\bar p \delta H \bar n^i + \bar p \bar H \delta n^i\ .
\eqno(58)
$$
Regarding (53), (54) and (55) we obtain

$$
\delta n_i=(\ 0, -\delta R_{,\theta},\ \ 0)\ ,
\eqno(59)
$$
$$
\delta n^i=(0, -\bar R^{-2}\delta R_{,\theta},\ 0)\ ,
\eqno(60)
$$
$$
\delta H = -{1\over\bar R^2}\delta R_{,\theta\theta}+ \cot\theta( -{1\over
\bar R^2}\delta R_{,\theta}) -{2\over\bar R^2}\delta R\ .
\eqno(61)
$$
The linearized equations of motion (35) can thus be written in the form
$$
\bar\sigma\delta A^i=-\delta^i_\theta\bar R^{-2}\delta p_{,\theta}
+\bar H \bar n^i\delta p +\bar p \bar n^i\delta H +\bar p\bar H \delta n^i
$$
$$
-{1\over 2}\delta\sigma\bar g^{ij}(^+\bar\Phi_{,j}+^-\!\bar\Phi_{,j})
-{1\over 2}\bar\sigma\bar g^{ij}\delta(^+\Phi_{,j}+^-\!\Phi_{,j})\ ,
\eqno(62)
$$
where $i,j=r,\theta$, $\bar g^{rr}=1$, $\bar g^{\theta\theta}=\bar R^{-2}$ (and
we do not need the $\varphi$--components because of axial symmetry).

Before calculating the perturbations of the gravitational potential let us
make the standard assumption that all quantities can be decomposed into
spherical harmonics. Thus, we write
$$
\delta\sigma=\sum_{l=0}^{\infty}\delta\sigma_l(t)Y_l\ ,\ \ \
\delta p=\sum_{l=0}^{\infty}\delta p_l(t)Y_l\ ,
\eqno(63)
$$
and
$$
\delta R=\sum_{l=0}^{\infty}\xi_l(t)Y_l\ ,\ \ \ \
\delta W=\sum_{l=0}^{\infty}\dot\eta_l(t)Y_{l,\theta}\ ,
\eqno(64)
$$
where $Y_l=Y_{l0}(\theta)$, $Y_{l,\theta}={\partial
Y_{l0}/\partial\theta}$, the form of $\delta R$ and $\delta W$ corresponds to the
fact that $\delta R$ describes a shift whereas $\delta W$ is a $\theta$--component
of a velocity (cf. eq. (46)). Because $Y_{l,\theta\theta}+\cot\theta\
Y_{l,\theta}=-l(l+1)Y_l$, we obtain from equation (61)

$$
\delta H=\sum_{l=0}^{\infty}\delta H_lY_l\ ,\ \
\delta H_l =\left(-{2\over\bar  R^2}+{l(l+1)\over\bar R^2}\right)\xi_l\
Y_l\ .
\eqno(65)
$$

The perturbations of the potential can be calculated by integrating the
Poisson equation (36) with $\rho$ obtained by perturbing the
$\delta$--function source (37). We shall proceed somewhat differently but we
checked that both procedures lead to the same result. Decompose the potential
inside and outside the shell into spherical harmonics (we are now omitting the
argument $t$ since it is irrelevant here):
$$
^-\Phi=\sum_{l=0}^\infty a_l(\epsilon)r^lY_l\ ,\ \ \ \ \  ^+\Phi=\sum_{l=0}^\infty
b_l(\epsilon)r^{-l-1}Y_l\ ,
\eqno(66)
$$
$$
^-\Phi_{,r}=\sum_{l=0}^\infty a_l(\epsilon)lr^{l-1}Y_l\ ,\ \ \
^+\Phi_{,r}=\sum_{l=0}^\infty b_l(\epsilon)(-l-1)r^{-l-2}Y_l\ .
\eqno(67)
$$
At the shell, $r=R(\theta,\epsilon)$, the potential is continuous,
$$
^-\Phi[R(\theta,\epsilon),\epsilon]=^+\Phi[R(\theta,\epsilon),\epsilon]\ ,
\eqno(68)
$$
and its gradient satisfies
$$
\left\{n^i[^+\Phi(r,\theta,\epsilon)_{,i}-^-\Phi(r,\theta,\epsilon)_{,i}]\right\}_{r=R(\theta,\epsilon)}
=4\pi G\sigma(\theta,\epsilon)\ .
\eqno(69)
$$
Linearization of these relations  with $\delta R=\sum_{l=0}^\infty \xi_l Y_l$ ,
$R(\theta,0)=\bar R$,  and
$\bar a_l=0$, $\bar b_l=0$ for $l\ge1$, implies
$$
\delta a_l\bar R^l=\delta b_l\bar R^{-l-1} -\bar b_0Y_0 \bar R^{-2} \xi_l\ ,
\eqno(70)
$$
and
$$
-(l+1)\delta b_l\bar R^{-l-2}+ 2\bar b_0Y_0 \bar R^{-3}\xi_l-l\delta a_l \bar R ^{l-1}=
4\pi G \delta\sigma_l\ ,
\eqno(71)
$$
where $Y_0={1/\sqrt{4\pi}}$.
We can solve for $\delta a_l$, $\delta b_l$ in terms of $\xi_l$ and $\delta\sigma_l$:
$$
\delta a_l={1\over 2l+1}\left(-(l-1)\bar b_0 Y_0\bar R^{-l-2} \xi_l-4\pi G \bar
R^{-l+1}\delta\sigma_l\right)\ ,
\eqno(72)
$$
$$
\delta b_l={1\over 2l+1}\left((l+2)\bar b_0Y_0\bar R^{l-1}\xi_l-4\pi G \bar R^{l+2}
\delta\sigma_l\right)\ .
\eqno(73)
$$
For the static background shell we have $\bar b_0Y_0=-4 \pi G \bar R^2 \bar\sigma$. Using this we
obtain
$$
\delta a_l={4\pi G\over 2l+1}\left[(l-1)\bar\sigma \xi_l- \bar
R\delta\sigma_l\right]\bar R^{-l}\ ,
\eqno(74)
$$
$$
\delta b_l={4\pi G\over 2l+1}\left[-(l+2)\bar\sigma\xi_l- \bar R
\delta\sigma_l\right]\bar R^{l+1}\ .
\eqno(75)
$$
The gravitational intensity at the shell is
$$
{\sl F}_i=-{1\over 2}\left[^+\Phi(r,\theta,\epsilon)_{,i}+^-\Phi(r,\theta,\epsilon)_{,i}
\right]_{r=R(\theta,\epsilon)}\ .
\eqno(76)
$$
Inserting $^\pm\Phi$ we obtain after linearisation, using that the background is spherically
symmetric, for the covariant $\theta$--component (with fixed angular behavior $Y_l$) of the
force intensity at the shell
$$
{\sl F}_{l\theta}=-{1\over 2}\left[\delta a_l\bar R^l +\delta b_l \bar R^{-l-1}\right]Y_{l,\theta}\ ,
\eqno(77)
$$
and the radial component is
$$
{\sl F}_{lr}=-{1\over 2}\left[\delta a_ll\bar R^{l-1} -\delta b_l(l+1) \bar R^{-l-2}
+2\bar b_0Y_0 \bar R ^{-3} \xi_l \right]Y_l\ .
\eqno(78)
$$
Inserting $\delta a_l, \delta b_l$ and $\bar b_0$ in which for $l=0$ we omit the background term,
${-GM/ 2\bar R^2}$, since it drops out as a consequence of equation (44),
we obtain
$$
{\sl F}_{l\theta}=-{1\over 2}{4\pi G\over 2l+1}\left[-3\bar\sigma\xi_l-2\bar R
\delta\sigma_l\right]Y_{l,\theta}\ ,
\eqno(79)
$$
and
$$
{\sl F}_{lr}=-{1\over 2}{4\pi G\over 2l+1}\left[\{l(l-1)+(l+1)(l+2)\}\bar R^{-1}\bar\sigma\xi_l+
\delta\sigma_l\right]Y_l
$$
$$+4\pi G\bar\sigma\bar R^{-1}\xi_l Y_l\ .
\eqno(80)
$$

It remains to consider the linearized form of the continuity equation (24). Since
$\bar\sigma=$const, and
$\bar V^\alpha=\bar U^j=0$, the first term, ${d\sigma/ dt}={\partial\sigma/\partial t}+
V^\alpha{\partial\sigma/\partial y^\alpha}$, after linearisation just becomes
$\delta\dot\sigma={\partial\delta\sigma/\partial t}$, the second term becomes
$\bar\sigma(\delta W_{,\theta}+\delta W\cot\theta)$, and the third $\bar\sigma\bar n_r\delta\dot R\bar
H$. Substituting the angular decompositions (63), (64), and using
$Y_{l,\theta\theta}+\cot\theta\  Y_{l,\theta}=-l(l+1)Y_l$, we obtain the $l$--part of the
continuity equation in the form
$$
\delta\dot\sigma_l-l(l+1)\bar\sigma\dot\eta_l+2\bar\sigma\bar R^{-1}\dot\xi_l=0\ .
\eqno(81)
$$
Integrating we get
$$
\delta\sigma_l-l(l+1)\bar\sigma\eta_l+2\bar\sigma\bar R^{-1}\xi_l=0\ ,
\eqno(82)
$$
where we put the integration constants equal to zero since $\xi_l=\eta_l=0$ implies $\delta\sigma_l=0$.

Multiplying now the intensity components (79), (80) by $\bar\sigma$, assuming the
background conditions (44) satisfied, and substituting the perturbed quantities as given
above into the equations of motion (62), we obtain
$$
\bar\sigma\ddot\xi_l=2 \bar R^{-1}\delta p_l+{1\over 4}G\bar\sigma M\bar R^{-3}[l(l+1)-2]\xi_l-
{1\over 2}GM\bar R^{-2}\delta\sigma_l
$$
$$
-{1\over 2}\bar\sigma{4\pi G\over 2l+1}[2l(l-1)\bar\sigma\bar R^{-1}\xi_l + \delta\sigma_l]\ ,
\eqno(83)
$$
$$
\bar\sigma\ddot\eta_l=- \bar R^{-2}\delta p_l- {1\over 2}G\bar\sigma M\bar R^{-4}\xi_l
+{1\over 2}\bar\sigma{4\pi G\over 2l+1}\bar R^{-2}(3\bar\sigma\xi_l +2\bar R \delta\sigma_l)\ ,
\eqno(84)
$$
where the first equation is meaningful for all $l\ge0$, the second for $l\ge 1$.

Finally, let us assume that the perturbed pressure and matter densities are connected by a linear relation
$\delta p=\alpha\delta\sigma$, so that
$$
\delta p_l=\alpha\delta\sigma_l\ .
\eqno(85)
$$
This, by $p=p(\sigma)$, $\delta p=({dp/ d\sigma})\delta\sigma$, corresponds to a general equation of
state for  barotropic fluids. Substituting for $\delta p$ into (83),(84), writing
$\bar\sigma={M/4\pi\bar R ^2}$, and expressing $\delta\sigma_l$ from the continuity equation
(68), we arrive at a system of two equations for just $\xi_l$ and $\eta_l$ as follows:
$$
\ddot\xi_l={GM\over\bar  R^3}\left[
-{l^2 -3l -2\over 2l+1}-{2-l(l+1)\over 4}-{4\alpha\bar R\over GM}
\right]\xi_l
$$
$$
+{GM\over\bar  R^2}\left[
-{l(l+1)^2\over 2l+1}+2l(l+1){\alpha\bar R\over GM}
\right]\eta_l\ ,
\eqno(86)
$$
$$
\ddot\eta_l={GM\over\bar  R^4}\left[
-{l+1\over 2l+1}+{2\alpha\bar R\over GM}
\right]\xi_l
+{GM\over\bar  R^3}\left[
{l(l+1)\over 2l+1}-l(l+1){\alpha\bar R\over GM}
\right]\eta_l\ ,
\eqno(87)
$$
where for $l=0$ only the first equation is meaningful.
Since we already used the continuity equation the last
two equations are the only equations to be solved for $\xi_l$, $\eta_l$ to determine the general
axisymmetric linearized perturbations.
\vskip 2cm
{\bf  5. The stability analysis }
\vskip 1cm
Before investigating stability let us cast the last coupled equations (86), (87)
into a still simpler form. Denoting
$$
\tilde\xi_l=\xi_l\ ,\ \ \  \ \ \tilde\eta_l=\bar R\eta_l\ ,\ \ \ \ \ \beta=\alpha{\bar
R\over GM}\ ,
\eqno(88)
$$
and the dimensionless time coordinate
$$
\tau=\left({GM\over \bar R^3}\right)^{1\over 2}t\ ,
\eqno(89)
$$
we get
$$
{d^2\tilde\xi_l\over d\tau^2}+A\tilde\xi_l+B\tilde\eta_l=0\ ,
\eqno(90)
$$
$$
{d^2\tilde\eta_l\over d\tau^2}+C\tilde\xi_l+D\tilde\eta_l=0\ ,
\eqno(91)
$$
where the coefficients are
$$
A={l^2 -3l -2\over 2l+1}+{2-l(l+1)\over 4}+4\beta\ ,
$$
$$
B={l(l+1)^2\over 2l+1}-2l(l+1)\beta\ ,
$$
$$
C={l+1\over 2l+1}-2\beta\ ,
$$
$$
D=-{l(l+1)\over 2l+1}+l(l+1)\beta\ .
\eqno(92)
$$

Applying $d^2/d\tau^2$ to equation (90) and regarding equation (91), we obtain the
following 4--th order equation for $\tilde\xi_l$,
$$
{d^4\tilde\xi_l\over d\tau^4}+(A+D){d^2\tilde\xi_l\over
d\tau^2}+(AD-BC)\tilde\xi_l=0\ ,
\eqno(93)
$$
and the same equation for $\tilde\eta_l$. Assuming
$\tilde\xi_l=\Xi_le^{i\omega{_l}\tau}$, the last equation implies
$$
\omega^4_l -(A+D)\omega_l^2 + AD-BC=0\ .
\eqno(94)
$$
Hence, the frequencies of the oscillations are given by
$$
\omega_l^{(1,2)}=\pm\left\{{1\over 2}(A+D)
\pm\left[{1\over 4}(A+D)^2-(AD-BC)\right]^{1\over 2}
\right\}^{1\over 2}\ ,
\eqno(95)
$$
where $(1,2)$ refers to the $\pm$ sign inside the bracket; the first sign (outside
the bracket) represents a trivial alternative.
The system is stable if --- given a fixed $\beta$ --- the $\omega_l$'s are real for
all $l$.

Let us first look at radial oscillations. With $l=0$ we have $A=-{3\over 2}+4 \beta$,
 $B=0$,
$C=1-2\beta$, $D=0$, so that the expression (95) gives
$$
\omega_{l=0}=\pm 2\left(\beta-{3\over 8}\right)^{\!\!{1\over2}}\ .
\eqno(96)
$$
The second solution of the relation (94), $\omega=0$, has no meaning since for $l=0$
only the first equation (92) with $\tilde\eta=0$ is valid. Therefore, we conclude
that {\it the shell is stable with respect to radial oscillations if }
$$
\beta>{3\over 8}.
$$
Using equation of state $\delta p  = \alpha\ \delta\sigma$ and relations (44), (88), we
can write this as the condition
$$
 \alpha = {\delta p \over \delta\sigma} > {3GM\over 8\bar R} \ \ \ \ \Leftrightarrow \ \ \ \ \
{\bar\sigma\over\bar p} {\delta p \over \delta\sigma} > {3\over 2}.
\eqno(97)
$$
This displays the analogy to the standard stability condition,
$\Gamma_1 = (\rho/p) (\delta p/\delta\rho) > 4/3$, for radial adiabatic stellar oscillations.
For stronger gravity (${3GM/8\bar R\ }$ large),   stiffer  equation of state,
$\delta p=\alpha\ \delta\sigma$, is needed to guarantee stability.

Turning next to dipole ($l=1$) perturbations we get $A=-{4\over3} +4\beta=-B$,
$C={2\over3}-2\beta=-D$, and the formula (95) yields
$$
\omega_{l=1}^{(1)}=0\ ,\ \ \ \ \ \omega_{l=1}^{(2)}=2(3\beta-1)^{1/2}\ .
\eqno(98)
$$
Regarding the equations (90), (91), we easily find out that the second solution
implies a trivial amplitude $\tilde\xi_1=\tilde\eta_1=0$, whereas the first,
$\omega_{l=1}^{(1)}=0$, implies time independent amplitudes; by incorporating the
angular parts we easily see that, as usually, they just describe a (small) shift of the
origin of the coordinates along the axis $\theta=0,\pi$.

For quadrupole ($l=2)$ perturbations we get
$$
{1\over2}(A+D)=-{3\over 2}+5\beta\ ,\ \ \ \ AD-BC=-{6\over 5}\beta\ ,
\eqno(99)
$$
so that the equation for the frequencies, (95), implies
$$
\omega_{l=2}^{(1)}=\pm\left\{-{3\over 2}+5\beta
+\left[(-{3\over 2}+5\beta)^2+{6\over5}\beta\right]^{1\over 2}
\right\}^{1\over 2}
\eqno(100)
$$
$$
\omega_{l=2}^{(2)}=\pm\left\{-{3\over 2}+5\beta
-\left[(-{3\over 2}+5\beta)^2+{6\over5}\beta\right]^{1\over 2}
\right\}^{1\over 2}\ .
\eqno(101)
$$
Since for stable radial oscillations we must have $\beta>{3\over8}$,\ \
  $\omega_{l=2}^{(2)}$ is not real and therefore the {\it self--gravitating fluid shell is
unstable} with respect to quadrupole pertubations.

Although this result is of course sufficient to prove instability let us look at
perturbations with large $l$. We find equations (92) to imply for $l\to\infty$
$$
{1\over 2}(A+D)\rightarrow{1\over
2}l^2(\beta-{1\over4})+{1\over2}l(\beta+{5\over4})+O(1)\ ,
$$
$$
AD-BC\rightarrow-{1\over4}l^2\beta +O(l)\ .
\eqno(102)
$$
Again, $\beta>{3\over8}$ implies ${1\over2}(A+D)>0$ but $AD-BC$ is negative so that
$\omega^{(2)}_{l\to\infty}$ is imaginary --- there is an instability.

Let us now turn to the amusing case of shells with negative mass $M$ (and with also
$\bar\sigma<0$ as the inertial mass desity). Putting first $l=0$ in the original
equation (86) for $\xi_l$, and defining $\beta$ again by (88), we obtain
$$
\ddot\xi_0+{GM\over\bar R^3}\left(-{3\over2}+4\beta\right)\xi_0=0\ .
\eqno(103)
$$
Introducing dimensionless time
$$
\tau=\left({-GM\over \bar R^3}\right)^{1\over 2}t\ ,
\eqno(104)
$$
we find
$$
\omega_{l=0}=\pm2\left(-\beta+{3\over 8}\right)^{\!\!{1\over2}}\ .
\eqno(105)
$$
Since $\beta=\alpha\bar R/GM$, we get stability for any $\beta$ such that
$$
-\infty<\beta<{3\over8}\ ,\ \ \  {\rm i.e.}\ \  -{3G(-M)\over 8\bar R}<\alpha<\infty\ .
\eqno(106)
$$

Introducing $\tau$ from equation (104) into the expressions (86), (87), we obtain
again the equations (90), (91), only with negative signs of $A,B,C,D$ given by
(92). Thus, the frequencies of oscillations of the shells with $M<0$ are
$$
\omega_l^{(1,2)}=\pm\left\{-{1\over 2}(A+D)
\pm\left[{1\over 4}(A+D)^2-(AD-BC)\right]^{1\over 2}
\right\}^{1\over 2}\ .
\eqno(107)
$$
The difference between this expression and expression (95), i.e. the sign change at
the first term ${1\over2}(A+D)$, is crucial. For dipole perturbations nothing new
arises --- one solution for $\omega$ leads to  vanishing amplitudes, the other
represents just a shift of the origin. However, turning to the quadrupole oscillations
we now get frequencies
$$
\omega_{l=2}^{(1.2)}=\pm\left\{{3\over 2}-5\beta
\pm\left[({3\over 2}-5\beta)^2+{6\over5}\beta\right]^{1\over 2}
\right\}^{1\over 2}\ ,
\eqno(108)
$$
which are all real provided that
$$
\beta<0\ ,\ \ \ {\rm i.e.}\ \ \ \alpha>0\ .
\eqno(109)
$$
Similarly, regarding the expressions (102) for large $l$, and taking $\beta<0$ we
find all frequencies real:
$$
\omega^{(1)}_{l\to\infty}\simeq\pm{1\over2}l(1-4\beta)^{1\over2}\ ,\ \ \ \ \ \ \
\omega^{(2)}_{l\to\infty}\simeq\pm{\beta\over1-4\beta}\ .
\eqno(110)
$$
In  fact, one can prove that for  {\it  any}   integer $l>1$  the expression  in superbrackets
under the
square root  in equation  (107) is positive. Indeed, the expression  under the square root
in the
square  brackets is a parabola in $\beta$   for any fixed  $l$ and one can show
that its minimum  is
always positive.

Next, one finds that
$$
-{1\over 2} (A+D)=-{1\over 2}(l^2+l+4) \beta + {1\over 8}(l^2+l+6)
$$
which is positive for $\beta<0$. Hence the expression  with the plus sign  in the
superbrackets is positive.
By forming its product  with the corresponding  expression with the minus sign  one obtains
$$
AD-BC ={l(l-1)(l+1)\over 4(2l+1)} [-\beta (2l-3)(l+2)+l-2]
$$
 which is always positive for integer $l>1$
(though  not e.g.  for $l=1.2$), and so $\omega_l^{(1,2)}$ is always real.

Thus, we conclude that {\it self--gravitating spherical shells with negative both
inertial and gravitational mass are stable} with respect to small oscillations.

Finally, it is of interest to consider an analogous case in electrostatics. We thus
neglect all inductive and radiative properties of the electromagnetic field, taking
only into account the fact that the same charges repel each other by Coulomb's law.
This should be a reasonable approximation for charged bubbles. Hence, we put $M=Q>0$,
the total charge of the sphere, and consider a negative gravitational constant
$-G=\gamma>0$. In fact, we can take any negative value for $\gamma$, multiply by
$\gamma^{-1}$ both original equations (83) and (84), and identify then by
$$
\bar\Sigma={\bar\sigma\over\gamma}\ ,\ \ \ \delta\Sigma={\delta\sigma\over\gamma}
\eqno(111)
$$
the inertial surface mass density and its perturbation, whereas leaving
$\bar\sigma=Q/4\pi \bar R^2$ and $\delta\sigma$ as the background charge density and
its perturbation. (We thus consider a fluid of charged particles with a fixed value
of the specific charge.)

First, from the equilibrium condition (43) we find  that in order to have a static
equilibrium we need a negative pressure, i.e. a tension
$$
\bar p =-\gamma{Q^2\over14\bar R^4}
\eqno(112)
$$
which balances the repulsive electric force.
Next we assume again a linear relationship
$$
\delta p_l=\alpha\delta\sigma_l=\alpha\gamma\delta\Sigma_l\ .
\eqno(113)
$$
The equation of continuity, (82), is valid for both inertial mass and charge.
However, in the terms on the r.h.s. of the equations of motion (83), (84), giving
the electric force, we of course have to substitute $\delta\sigma_l$.  It is easy to
see that the equations (86) and (87) become
$$
\gamma^{-1}\ddot\xi_l=-{Q\over\bar  R^3}\left[
-{l^2 -3l -2\over 2l+1}-{2-l(l+1)\over 4}+{4\alpha\bar R\over\gamma Q}
\right]\xi_l
$$
$$
-{Q\over\bar  R^2}\left[
-{l(l+1)^2\over 2l+1}-2l(l+1){\alpha\bar R\over\gamma Q}
\right]\eta_l\ ,
\eqno(114)
$$
$$
\gamma^{-1}\ddot\eta_l=-{Q\over\bar  R^4}\left[
-{l+1\over 2l+1}-{2\alpha\bar R\over\gamma Q}
\right]\xi_l
-{Q\over\bar  R^3}\left[
{l(l+1)\over 2l+1}+l(l+1){\alpha\bar R\over\gamma Q}
\right]\eta_l\ .
\eqno115)
$$
Therefore, writing $\tilde\xi_l=\xi_l$, $\tilde\eta_l=\bar R\eta_l $ as in
equation (88) and defining now
$$
\beta=-{\alpha\bar R\over\gamma Q}=+{\alpha\bar R\over G Q}\ ,
\eqno(116)
$$
and
$$
\tau=\left({\gamma Q\over \bar R^3}\right)^{1\over 2}t\ ,
\eqno(117)
$$
we arrive at
$$
{d^2\tilde\xi_l\over d\tau^2}-A\tilde\xi_l-B\tilde\eta_l=0\ ,
\eqno(118)
$$
$$
{d^2\tilde\eta_l\over d\tau^2}-C\tilde\xi_l-D\tilde\eta_l=0\ ,
\eqno(119)
$$
where $A,B,C,D$ are given again by  equation (92).
These are exactly the same equations as those we analyzed for the shells with
negative mass. Consequently, we can conclude that the charged shells will be stable
with respect to radial and non--radial oscillations if the parameter $\beta$ in
equation (116) is negative. Since $G<0$, this requires
$$
\alpha > 0\ ,
\eqno(120)
$$
i.e., regarding equation (113), a decrease of the tension when the mass and the
charge density increase. Consider a static charged shell in which the repulsive
effects of the charges is compensated by a tension. If the radius of the shell is
increased while its total charge is unchanged, the repulsion decreases and a decreased
tension can still bring back the shell into the original radius.

The results of the stability analysis of all three problems we discussed can, in
fact, be made intuitively plausible by considering the following Figure 1.


In Figure 1a a perturbed self--gravitating shell with positive mass is illustrated.
Gravity is pointing always inwards. But the perturbed element is pushed inwards also by
the surface pressure exerted by adjacent elements, so an instability arises. In Figure
1b a shell with a negative mass is considered. Gravity now acts outwards, the
pressure acts also outwards but it causes an acceleration pointing inwards because
the inertial mass of the element is negative (see the discussion at the end of
Section 3), and thus the shell with negative mass can be in stable equilibrium. If
the element is pushed inwards as in Figure 1a, both gravity and pressure give it an
acceleration pointing outwards. With charged shells the situation is similar to that
of a shell with a negative mass. It is now a tension which pulls an element inwards
in Figure 1b and thus acts against the electric repulsion.

In the theory of adiabatic nonradial stellar oscilations the Schwarzschild discriminant,
$A = (d\ln\rho/dr) - \Gamma_1^{-1} (d\ln p/dr)$, plays an important role. If the criterion
of convective stability, $A < 0$, is violated in the whole star, unstable modes exist
(e.g. Ledoux \& Walraven 1958). This corresponds to our considerations illustrated in
Figure 1. In the situation described in Figure 1a the element suffers a ``convective
instability'', whereas it is stable in the situation depicted in Figure 1b.

\bigskip
\bigskip

\vskip 1cm
{\bf Acknowledgment:}

We thank Douglas Gough from the Institute of Astrophysics, Cambridge, for saying that
self--gravitating fluids are probably unstable but have apparently not been treated
in the literature. We are grateful to J\"urgen Ehlers for reading the manuscript and making
helpful suggestions. We also acknowledge discussions with Peter H\"ubner, Jerzy Lewandowski,
Jim Pringle, and Allan Rendall. We  are thankful to Vojt\v ech  Pravda for demonstrating
the
reality of expression (107) and help with the manuscript. J.B. is grateful to the Albert
Einstein Institute for kind hospitality, and to the grants GACR--202/96/0206 and
GAUK--230/1996 of the Czech Republic and the Charles University for a partial support.
\vfill\eject
{\bf  References}
\vskip 1cm

Aris, R. 1989, Vectors, Tensors, and the Basic Equations of Fluid Mechanics (New
York: Dover)

Barrab\`es, C., \& Israel, W. 1991, Phys. Rev. D, 43, 1129

Berezin, V. A., Kuzmin, V. A., \& Tkachev, I. I. 1987, Phys. Rev. D, 36, 2919

Bi\v c\'ak, J., \& Ledvinka, T. 1993, Phys. Rev. Lett., 71, 1669

Blau, S. K., Guendelman, E. I., \& Guth, A. H. 1987, Phys. Rev. D, 35, 1747

Bondi, H. 1957, Rev. Mod. Phys. 29, 423

Brady, P. R., Louko, J., \& Poisson, E. 1991, Phys. Rev. D, 41, 1891

Friedman, J. L., Louko, J., \& Winters--Hilt, S. N. 1997, Phys. Rev. D. 56, 7674

H\'aj\' \i\v cek, P., \& Bi\v c\' ak, J. 1997, Phys. Rev. D, 56, 4706

Kellog, O. D. 1967, Foundations of Potential Theory (Berlin: Springer--Verlag)

Kolotch, S. J., \& Eardly, D. M. 1997, Phys. Rev. D, 56, 4651 and 4663

Ledoux, P., \& Walraven, Th. 1958, Variable Stars, in Handbuch der Physik 51, ed. S. Fl\"ugge
(Berlin: Springer-Verlag)

Love, A. E. H. 1944, A Treatise on Mathematical Theory of Elasticity, 4th edition (New
York: Dover)

Marsden J. E., \& Hughes, T. J. R. 1983, Mathematical Foundations of Elasticity (Englewood
Cliffs: Prentice--Hall)

Naghdi, P. M. 1972, The Theory of Shells, in Handbuch der Physik VIa/2, ed. C. Truesdell
(Berlin: Springer--Verlag)

N\' u\~ nez, D. 1997, Ap. J., 482, 963

Peebles, P. J. E. 1993, Principles of Physical Cosmology  (Princeton: Princeton University
Press)

Price, R. H. 1993, Amer. J. Phys., 61, 216

Purcell, E. M. 1965, Electricity and Magnetism --- Berkely Physics Course, Vol. 2 (New
York: McGraw--Hill)

Sato, H. \& Yamada, Y. 1991, Prog. Theor. Phys., 85, 541

Schmidt, B. 1999, Phys. Rev. D59, 024005

Scriven, L.E. 1960, Chem. Engng. Sci., 12, 98

Turok, N. ed. 1997, Critical Dialogues in Cosmology (Singapore: World Scientific)

Vishniac, E. T. 1983, Ap. J., 247, 152

Wald, R. M. 1984, General Relativity (Chicago: The University of Chicago Press)

\vfill\eject
$$\vbox{
\settabs 2 \columns
\+ \epsfbox{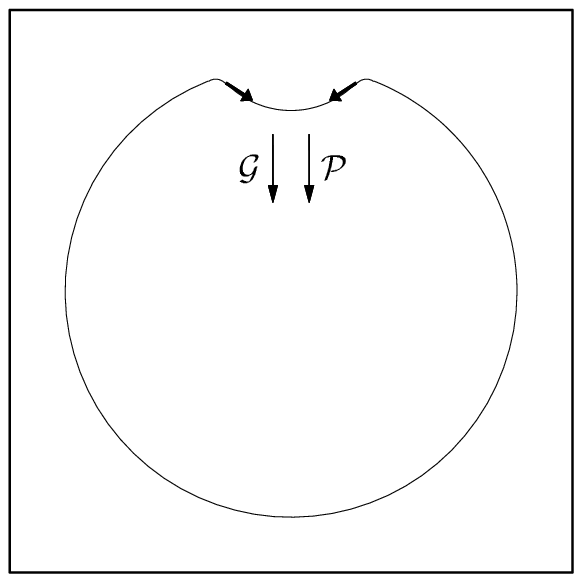} & \epsfbox{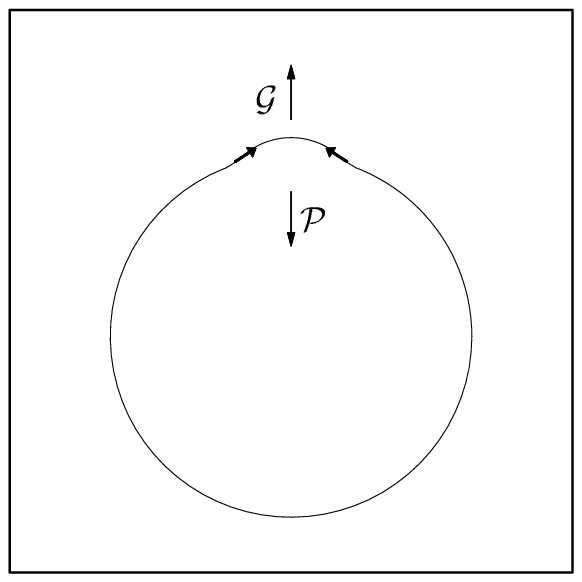} \cr
}$$

Figure 1
\hfil\break
a) A perturbed self-gravitating shell with positive mass. Both gravity and pressure
on the perturbed element point inwards. The shell is unstable. \hfil\break
b) A perturbed self-gravitating shell with negative gravitational and inertial mass.
Both gravity and pressure on the perturbed element point outwards but the acceleration caused
by the pressure points inwards due to the negative inertial mass. A similar situation
arises with charged shells with tension. Such shells are stable.

\end